\documentclass[12pt]{article}

\usepackage{amsmath,amssymb}
\usepackage{cite}
\usepackage{graphicx}

\begin{document}
\begin{titlepage}
\begin{center}

 \vspace{-0.7in}

{\large \bf  Bubble nucleation in disordered Landau-Ginzburg model}\\
%\vspace{.06in}
% and\\\vspace{.18in}
% Interfaces in Random Media}\\
\vspace{.3in}
{\large\em R. Acosta Diaz\,\,\footnotemark[1] and N. F. Svaiter\,\,\footnotemark[2]}\\
\vspace{.08in}

Centro Brasileiro de Pesquisas F\'\i sicas - CBPF\\
Rua Dr. Xavier Sigaud, 150, Rio de Janeiro - RJ, 22290-180, Brazil\\

\vspace{.3in}

{\large\em C.A.D. Zarro\,\,\footnotemark[3]}\\
\vspace{.08in}

Universidade Federal do Rio de Janeiro, Instituto de F\'isica\\
Av. Athos da Silveira Ramos, 149, Rio de Janeiro - RJ, 21941-909, Brazil

\vspace{.3in}

\subsection*{\\Abstract}
\end{center}

In this paper we investigate bubble nucleation in a disordered Landau-Ginzburg model. First we adopt the standard 
procedure to average over the disordered free energy. This quantity is represented as a series of the replica 
partition functions of the system. Using the saddle-point equations in each replica partition function, we discuss 
the presence of a spontaneous symmetry breaking mechanism. The leading term of the series is given by a large-$N$ 
Euclidean replica field theory. Next, we consider finite temperature effects. Below some critical temperature, there 
are $N$ real instantons-like solutions in the model. The transition from the false to the true vacuum for each 
replica field is given by the nucleation of a bubble of the true vacuum. In order to describe these irreversible 
processes of multiple nucleation, going beyond the diluted instanton approximation, an effective model is 
constructed, with one single mode of a bosonic field interacting with a reservoir of $N$ identical two-level systems. 

\bigskip
\vspace{.06in}

\noindent
{\sc keywords:} disordered systems; free energy; bubble nucleation.
\vspace{.06in}

\footnotetext[1]{e-mail:\,\,racosta@cbpf.br}
\footnotetext[2]{e-mail:\,\,nfuxsvai@cbpf.br}
\footnotetext[3]{e-mail:\,\,carlos.zarro@if.ufrj.br}
\vspace{.09in}
\noindent
PACS numbers: 05.20.-y,\,75.10.Nr

\end{titlepage}
\newpage\baselineskip .18in

\section{Introduction}\label{intro}
\quad

The critical behavior of disordered systems has been discussed since the $70'$s in the literature. Two concepts that 
are of fundamental importance in such systems are respectively frustration and quenched disorder. Frustration was 
introduced to describe properties of spin-glasses with many different ground states  \cite{and}. The free energy 
landscape of these systems have a multivalley structure. In quenched disordered systems, the disorder is in static 
equilibrium and therefore these systems are spatially random. The study of quenched disordered systems leads to new 
universality classes in critical regions and also the possibility of a large number of metastable states in free 
energy landscape. In such systems defined in the continuum with quenched disorder, it is a hard task to perform a 
perturbative expansion in any model, since these systems are intrinsically inhomogeneous.  One way to circumvent such 
problem is to average over the ensemble of all realizations of the disorder quantities of interest. For example,  
average the free energy functional with respect to the probability distribution of the disordered field. In these 
disordered systems, the replica symmetry breaking with its physical consequences, has been intensely discussed by the 
physical community \cite{edwards,SerKir,RBS1,RBS2,livro1,livro3,livro4}. 

Recently, it was proposed a new method to average the disorder dependent free energy \cite{distributional, 
distributional2}. Physical consequences of this approach were investigated in Refs. \cite{polymer,zarro}. The 
motivation of this paper are the following. First is to stress the main differences between perturbative expansions 
in field theories without or in the presence of disorder fields, discussing cluster properties of disordered average 
$n$-point correlation functions. The second one is to discuss the physical consequences of the results obtained in 
Ref. \cite{zarro}. Finally, going beyond the above discussed results, we introduce an effective model to describe 
false-true vacuum transitions of replica fields. Specifically, we are interested to describe the phase transitions 
present in the continuous version of the $d$-dimensional random field Ising model \cite{larkin, livro2, naterman}.  
Using the approach discussed above, the structure of the replica space is investigated using the saddle-point 
equations obtained from each replica field theory. Assuming the replica symmetric ansatz, we investigate the 
spontaneous symmetry breaking mechanism in some replica partition functions. Our approach reveals the existence of 
replica instantons-like solutions (real or complex) in this model \cite{dotsenko4,Maxin,Allan}. For the case of real 
instantons-like solutions our methodology produced the following scenario. 

Vacuum decay in this theory with $N$ replicas can be described in the following way. For low temperatures, there is a 
critical temperature where each replica field has two non-degenerate vacuum states. Consequently, for each replica 
field there will be a transition from the false vacuum to the true one with nucleation of a bubble of the true 
vacuum. This first-order phase transition, in the low temperature limit, was investigated in Refs. 
\cite{aharonytricritical,aharonyetal}. The crucial question here is which tools we can use to describe these 
irreversible processes, i.e., the nucleation of bubbles of true vacuum in the false vacuum environment.

We shall now be concerning with the description of the collective behavior of the $N$ replica fields. The point of 
departure is given by Ref. \cite{Leggett}. There, the authors emphasize that it is possible to represent $N$ 
structures with a false and a true vacuum using two-level systems. The situation where the nucleation of bubbles 
occurs, decreasing the free energy of the system is characteristic of an open system. To go further describing this 
multiple nucleation, i.e., the collective nucleation of bubbles in the disordered system, an effective model is 
constructed using the functional integral formalism developed to study phase transitions in quantum optics systems by 
Popov and Fedotov \cite{Popov1,Popov2,Popov3}.  In the large $N$ limit, the functional integral describes an ensemble 
of $N$ two-level systems interacting with one single bosonic mode, instead of the usual situation of a the countable 
infinitely set of field modes. 

The justifications for introducing the bosonic mode are the following: this bosonic mode is connecting the two-level 
systems and also it makes possible the decay for each replica field from the false  vacuum to the true one. In this 
scenario, it is possible to show the existence of a temperature where the free energy is non-analytic.
The equivalence between these two quite distinct physical models can be justified using the following argument. In 
the disordered Landau-Ginzburg model, the leading replica partition function in the series representation for the 
free energy shows that all the replica fields, with a false and true vacuum states are strongly correlated. This is 
exactly the situation discussed by Popov and Fetotov, where only one single mode is resonant with the two-level 
atoms. All the two-level atoms interact coherently with this single mode.

The organization of this paper is the following. In Sec. \ref{sec:disoderedLG}, we discuss a $d$-dimensional 
disordered Landau-Ginzburg model. In Sec. \ref{sec:distributionalzeta}, in a generic replica partition function we 
discuss  the structure of the replica space using the saddle-point equations of the model. In Sec. 
\ref{sec:replicainstantons}, we demonstrate at low temperatures the emergence of $N$ instantons-like solutions in the 
leading replica partition function of the model.  In Sec. \ref{sec:effectivemodel}, to describe the bubble nucleation 
in the disordered model, an effective model is constructed using the formalism developed by Popov and Fedotov. 
Conclusions are given in Sec. \ref{sec:conclusions}. We use the units $\hbar=c=k_{B}=1$.

\section{A disordered Landau-Ginzburg model}\label{sec:disoderedLG}
\quad

In magnetic materials with disorder, in principle there are two kind of systems. The first set is one where the 
disorder is related to the local spin interaction. In this case the disorder generates multiple disordered ground 
states, the spin-glass phase. The second, is one where the disorder is a random external perturbation. One disordered 
model that belongs to this second set is the random field Ising model. This model has been studied intensively from 
the theoretical and experimental point of view and used to describe many systems in nature. One is the case of 
diluted antiferromagnetic in a homogeneous external field \cite{anti1,anti2} and also binary fluids in porous media.
For instance, in order to model binary fluids confined in porous media, when the pore surfaces couple differently to 
the two components of a phase-separating mixture, the random field has been used by the literature. These systems can 
develop a second or a first-order phase transition \cite{fluids1,fluids2,fluids3}. The random field Ising model in a 
hypercubic lattice in $d$-dimensions is described by the Hamiltonian 

\begin{equation}
 H=-J\sum_{(i,j)}^{N}\,S_{i}S_{j}-\sum_{i}\,h_{i}S_{i}, \label{1}
\end{equation}
where $(i,j)$ indicates that the sum is performed over nearest neighbour pairs and $S_{i}=\pm 1$. In the above 
equation $N$ is the total number of Ising spins. Periodic boundary conditions can be used and the thermodynamic limit 
must be used in the end. The partition function is $Z=Tr\, e^{-\beta H}$. In  Eq. (\ref{1}) the $h_{i}$'s are the 
quenched random variables totally uncorrelated on different sites. The average free energy is defined by 
$F=-\frac{1}{\beta}\,\mathbb{E}[\ln Z]$, where $\mathbb{E}\,[...]$ means the average over the ensemble of all the 
realizations of the quenched disorder. Here  we consider a Gaussian distribution defined by

\begin{equation}
 P(h_{i})=\frac{1}{\sqrt{2\pi h_{0}^{2}}}\exp\bigg(\frac{-h_{i}^{2}}{2h_{0}^{2}}\biggr).  \label{no}
\end{equation}

The probability distribution of such quenched random variables has zero mean-value, $\mathbb{E}\,[h_{i}]=0$, and 
correlation functions given by $\mathbb{E}\,[h_{i}h_{j}]=h_{0}^{2}\delta_{ij}$. Here we are interested in the small 
disordered limit, i.e., $h_{0}^{2} << 1$.

The properties of the phase transition of the random field Ising model is still under debate 
\cite{mezard1,mezard2,dotsenko,orland,dotsenko2,dotsenko3,sherington1,sherington}. The question of the lower critical 
dimension, bellow which long-range order is absence and the upper critical dimension, above which the model presents 
mean-field behavior independent of the dimension has been a matter of controversy. Imry and Ma  obtained that the 
model with nearest neighbor interaction presents spontaneous magnetization only for $d\geq \,3$ \cite{ma}. This 
result is in contradiction with the dimensional reduction argument \cite{sourlas,parisi1}. The controversy was solved 
by Bricmont and Kupiainen, who proved that there is a phase transition in the random field Ising model for $d\geq\,3$ 
\cite{bric1,bric2}, and Aizenman and Wehr, that showed the absence of phase transition for $d=2$ in the 
model \cite{wehr1}. 

The behavior of systems defined in a lattice near the critical point can be modeled by continuous statistical field 
theories. This can be achieved replacing the lattice structure by a continuum where the order parameter  can be 
obtained averaging with respect to a statistical weight a random continuous field. For instance, the effective $O(n)$ 
Landau-Ginzburg model is defined by $\varphi_{i}(x)$, a $n$-component field. This model is able to describe several 
universality classes. For the case $n\rightarrow 0$ can describe self-avoided polymers \cite{gennes,gaspari,schafer}. 
For $n=1$ it describes the critical behavior of the Ising model. For $n=2$ the critical behavior of the $XY$ model 
and also the two-dimensional Coulomb gas are described. For $n=3$ the Heisenberg model \cite{livrozinnjustin} and low 
energy dynamics of QCD can be modeled by $n=4$. Finally for $n\rightarrow \infty$ it is possible to solve exactly the 
model. In this paper we study the critical properties of the random field Ising model, by means of a continuous 
scalar field theory defined in $\mathbb{R}^{d}$ with symmetry $Z_{2}$ (the $n=1$ case).

We assume that the critical behavior of the random field Ising model in $\mathbb{Z}^{d}$ can be described by a 
continuous disordered Landau-Ginzburg model. Firstly, let us briefly discuss the model without disorder. We are 
following Ref. \cite{JAC}. The Landau-Ginzburg functional, $i.e.$, the Hamiltonian $H(\varphi)$ for the scalar field 
is given by

\begin{equation}
 H=\int d^{d}x\Biggl(\frac{1}{2}\varphi(x)\left(-\Delta+m_{0}^{2}\right)\varphi(x)
                             +\frac{\lambda_{0}}{4!}\varphi^{4}(x)\Biggr),         \label{9}
\end{equation}
where the symbol $\Delta$ denotes the Laplacian in $\mathbb{R}^{d}$ and $\lambda_{0}$ and $m_{0}^{2}$ are analytic 
functions of the temperature. Actually, $m_{0}^{2}$ is the inverse of the correlation length. Doing a parallel with 
Euclidean field theory we call them respectively the bare coupling constant and the squared mass of the model. 
For high temperatures, away from the critical point, the correlation functions of the model are short-ranged. Near 
the critical point, the correlation functions becomes long ranged, where the characteristic length scale, the 
correlation length $\xi$ has a power law behavior, exactly as in Euclidean field theory. The partition function of 
the model is defined by the functional integral

\begin{align}
 Z=\int_{\partial\Omega} [d\varphi]\,\, &\exp\bigl(-H(\varphi)\bigr),  \label{88}
\end{align}
where $[d\varphi]$ is a formal Lebesgue measure, given by $[d\varphi]=\prod_{x} d\varphi(x)$, and $\partial\Omega$ in 
the functional integral means that the field $\varphi(x)$ satisfies some boundary condition in the boundary 
$\partial\Omega$ of some bounded domain, i.e., connected open set $\Omega \subset \mathbb{R}^{d}$. Periodic boundary 
conditions can be imposed to preserve translational invariance, replacing $\mathbb{R}^{d}$ by the torus 
$\mathbb{T}^{d}$. To remove the ultraviolet divergences, in the Fourier decomposition of the field a cut-off must be 
introduced. This cut-off is related with a elementary length scale, the lattice spacing of the original model. Since 
in all the discussions of this paper we need no more than the tree-level calculations, this technical remark is 
immaterial for the results presented in the paper.

The question now arises is the cluster properties of correlation functions for the model with or without disorder. 
Therefore, let us start briefly discussing these quantities. Averaging with respect to the Boltzmann weight we get 
the $n$-point correlation functions of the model 

\begin{equation}\label{eq:correlationpure}
\langle\varphi(x_{1})...\varphi(x_{n})\rangle=
           \frac{1}{Z}\int [d\varphi]\prod_{i=1}^{n}\varphi(x_{i}) \exp\bigl(-H(\varphi)\bigr).
\end{equation}
Introducing a fictitious source $j(x)$ we can define $Z(j)$, the generating functional of all $n$-point correlation 
functions as \cite{livro5,lk}

\begin{equation}
 Z(j)=\int_{\partial\Omega} [d\varphi]\,\, \exp\left(-H(\varphi)+\int d^{d}x\, j(x)\varphi(x)\right).
\label{eq:generatingfunctional}
\end{equation}
Taking functional derivatives with respect to the source and setting to zero in the end, we obtain the $n$-point 
correlations functions of the model

\begin{equation}\label{eq:correlationfunction2}
\langle\varphi(x_{1})..\varphi(x_{k})\rangle=
      Z^{-1}(j)\left.\frac{\delta^{k} Z(j)}{\delta j(x_{1})...\delta j(x_{k})}\right|_{j=0}. 
\end{equation}
Notice that these $n$-point correlation functions are given by the sum of all diagrams with $n$-external legs, 
including the disconnected ones. Next, using the linked cluster theorem, it is possible to define the generating 
functional of connected correlation functions given by $W(j)=\ln Z(j)$. The order parameter of the model without 
disorder $\langle\varphi(x)\rangle$ is given by

\begin{equation}
 \langle\varphi(x)\rangle=Z^{-1}(j)\left.\frac{\delta Z(j)}{\delta j(x)}\right|_{j=0}. 
\end{equation}

Before continue, we would like to clarify terminology. Although streaking speaking, the order parameter is defined by 
the above equation, throughout  this paper we may call local order parameter for the continuous field 
$\varphi(x)$ defined for $x \in \mathbb{R}^{d}$. Applying two functional derivatives on the generating functional of 
connected correlation functions we get 

\begin{align} \nonumber
 \langle\varphi(x_{1})\varphi(x_{2})\rangle_{connected} & =
      \biggl[\frac{1}{Z(j)}\frac{\delta^{2}Z(j)}{\delta j(x_{1})\delta j(x_{2})} \\
&\left.-\frac{1}{Z(j)^{2}}\frac{\delta Z(j)}{\delta j(x_{1})}\frac{\delta Z(j)}{\delta j(x_{2})}\biggr]\right|_{j=0}.
\end{align} 
The large distance decay properties of these connected correlation functions are called cluster properties. These 
correlation functions goes to zero for $|x_{1}-x_{2}|\rightarrow \infty$. In an Euclidean field theory the cluster 
properties of the Schwinger functions are equivalent to the uniqueness of the vacuum.

We briefly present the basic tools that we need to discuss disordered systems \cite{malivro}.  The continuum version 
for the $d$-dimensional random field Ising model, is given by a $d$-dimensional Landau-Ginzburg scalar 
$\lambda\varphi^{4}$ model in the presence of a disorder field linearly coupled to the scalar field. The Hamiltonian 
in the presence of disorder is given by

\begin{equation} 
 H(\varphi,h)=H(\varphi)+ \int d^{d}x\,h(x)\varphi(x),
\end{equation}
where  $H(\varphi)$ is the Landau-Ginzburg Hamiltonian, defined in Eq. (\ref{9}), and $h(x)$ is a quenched disorder 
field. The disordered functional integral $Z(h)$ is defined by 

\begin{equation}
 Z(h)=\int_{\partial\Omega} [d\varphi]\,\, \exp\bigl(-H(\varphi,h)\bigr).
\label{8}
\end{equation}
Eq. (\ref{8}) defines the partition function associated with the scalar field for a  given disorder configuration. 
The $n$-point correlation functions for one specific realization of the disorder field reads

\begin{equation}\label{eq:disordercorrelationfunction}
\langle\varphi(x_{1})..\varphi(x_{n})\rangle_{h}=\frac{1}{Z(h)}\int [d\varphi]\prod_{i=1}^{n}\varphi(x_{i}) 
\exp\bigl(-H(\varphi,h)\bigr).
\end{equation}

To introduce a generating functional for one realization of the disorder field, $Z(h;j)$, we again use a fictitious 
source $j(x)$:

\begin{equation}
 Z(h;j)=\int_{\partial\Omega} [d\varphi]\,\, \exp\biggl(-H(\varphi,h)+\int d^{d}x \,j(x)\varphi(x)\biggr).
\label{eq:disorderedgeneratingfunctional}
\end{equation}
For a particular realization of the disorder field, $Z(h;j)$ can be used to obtain the $n$-point correlation function 
given by  Eq. (\ref{eq:disordercorrelationfunction}) by means of functional derivatives. With these correlation 
functions, one can compute the disorder-averaged correlation functions given by 

\begin{equation}
\mathbb{E}\bigl[\langle\varphi(x_{1})...\varphi(x_{n})\rangle_{h}\bigr]=\int\,[dh]P(h)\langle\varphi(x_{1})...
\varphi(x_{n})\rangle_{h},
\end{equation} 
where $\langle\varphi(x_{1})..\varphi(x_{n})\rangle_{h}$ is given by Eq. (\ref{eq:disordercorrelationfunction}) and 
$[dh]=\prod_{x} dh(x)$ is again a formal Lebesgue measure. As in the pure system case, one can define a generating 
functional for one disorder realization, $W_{1}(h;j)=\ln Z(h;j)$. We take the disorder-average of this generating 
functional, $W_{2}(j)=\mathbb{E}[W_{1}(h;j)]$. We have 

\begin{equation}
 W_{2}(j)=\int\,[dh]P(h)\ln Z(h;j).
\label{eq:disorderedfreeenergy}
\end{equation} 
Taking the functional derivative of $W_{2}(j)$ with respect to $j(x)$, we get

\begin{equation}
\left.\frac{\delta W_{2}(j)}{\delta j(x)}\right|_{j=0}=\int\,[dh]P(h)\biggl[\frac{1}{Z(h;j)}\left.\frac{\delta 
Z(h;j)}{\delta j(x)}\biggr]\right|_{j=0}.
\end{equation}
Since $\langle\varphi(x)\rangle_{h}$ is the average of the field for a given configuration of the disorder in the 
disorderd Landau-Ginzburg model the above quantity $\mathbb{E}\bigl[\langle\varphi(x_{1})\rangle_{h}\bigr]$ is the 
order parameter of the model  \cite{parisi1}. The second functional derivative of $W_{2}(j)$ with respect to $j(x)$ 
gives $G(x_{1}-x_{2})$. We have

\begin{align}
\left.\frac{\delta^{2}W_{2}(j)}{\delta j(x_{1})\delta j(x_{2})}\right|
_{j=0}&=\mathbb{E}\left[\langle\varphi(x_{1})\varphi(x_{2})\rangle_{h}\right]\nonumber \\
& -\mathbb{E}\left[\langle\varphi(x_{1})\rangle_{h}\langle\varphi(x_{2})\rangle_{h}\right].
\label{eq:two-point}
\end{align}
Notice that, in general, the following quantities are not equal, i.e.,

\begin{equation}\label{eq:difference}
\mathbb{E}\left[\langle\varphi(x_{1})\rangle_{h}\langle\varphi(x_{2})\rangle_{h}\right]\neq 
\mathbb{E}\left[\langle\varphi(x_{1})\rangle_{h}\right]\mathbb{E}\left[\langle\varphi(x_{2})\rangle_{h}\right].
\end{equation}
Therefore the Eq. (\ref{eq:two-point}) is not the disordered average two-point connected correlation function. To 
proceed let us define the following averaged quantity 

\begin{equation}
 \chi(x_{1}-x_{2})=\mathbb{E}\left[\langle\varphi(x_{1})\rangle_{h}\langle\varphi(x_{2})\rangle_{h}\right].
\end{equation} 
This above disconnected correlation function can be different from zero even if the order parameter of the model is 
zero. The decay of these two-point correlation functions $G(x_{1}-x_{2})$ and $\chi(x_{1}-x_{2})$ at critical region 
defines two critical exponents $\eta$ and $\eta'$ \cite{rieger}. We have 

\begin{equation}
  G(x_{1}-x_{2})\approx |x_{1}-x_{2}|^{-(d-2+\eta)}.
\end{equation} 
and

\begin{equation}
 \chi(x_{1}-x_{2})\approx |x_{1}-x_{2}|^{-(d-4+\eta')}.
\end{equation} 

In a pure system, taking functional derivatives of $W(j)$ we get the connected correlation functions, that satisfies 
clustering property. Applying two functional derivatives, the disordered average functional $W_{2}(j)=\mathbb{E}
[W_{1}(h;j)]$ does not generate the disordered average two-point connected correlation functions of the model.
This can be generalized to the $n$-point correlation functions, or being more precise, investigating cluster 
properties of disordered average $n$-point correlation functions. The fundamental problem is the fact that since 
there are many minima \cite{Parisi88,aharony,fytas} in these systems, we can not expand around only one specific 
minimum, hence a non-perturbative scenario emerges. The non-perturbative scenario can not be studied neither using 
the renormalization group equations nor the composite operator formalism \cite{jackiw, ananos, gap}. Composite 
operator formalism is a way to use resummation methods (sum of infinite series of diagrams) to avoid the infrared 
divergences of a massless theory. These methods can not reveal the vacuum structure of the disordered system.  

One possible way to proceed is the following. In the presence of these metastable states one must identify clustering 
states, i.e., the states where the connected correlation functions vanishes for large distances, and introduce an 
order parameter that characterize such domain \cite{PRL}. We do not expect that this program can be implemented in a 
straightforward way. To deal with this above discussed problem, the first step is to identify the metastable states, 
i.e., show the presence of many local minima in the free energy landscape. In other words, this fundamental 
difficulty may point that a local approach of field theory based in the correlation functions must be substituted, at 
least in the beginning by another more promising procedure. Instead of concentrate our efforts to define local 
objects, we may study only global quantities, such as, the averaged free energy. As we expected, here we will show 
the presence of a large number of metastable states in the disordered system. 

For instance, for free fields without disorder the spectral zeta-function technique 
\cite{seeley,ray,hawking,dowker,fulling}, which is a way to regularize the determinant of Laplace operator, can be 
used to compute the free energy of this pure system. In the next section, we show how this approach can be used to 
access the non-perturbative landscape of the disordered system. Here, we proceed as follows. We are interested to 
compute $W_{2}(j)|_{j=0}=\mathbb{E}[W_{1}(h;j)]|_{j=0}$, namely the disorder-averaged free energy.

\section{Distributional zeta-function approach} \label{sec:distributionalzeta}
\quad

In order to circumvented the problem of many local minima that the perturbative expansion fail to take into account, 
Lancaster \emph{et al.} \cite{lancaster} discussed a model where many solutions of the mean field equations obtained 
from each realization of the disorder are weighted by Boltzmann factors. In the following we show that it is possible 
to investigate a non-perturbative scenario using the distributional zeta-function approach 
\cite{distributional,distributional2}. This approach has similarities with the above discussed method. 

Here, we do not give details of the derivation but only the essential steps of the mathematical rigorous procedure 
that allow to use the replica partition functions in order to compute the disorder-averaged free energy.  For a given 
probability distribution of the disorder, one is mainly interested in averaging the disorder dependent free energy 
functional which reads

\begin{equation}
 F=-\frac{1}{\beta}\int\,[dh]P(h)\ln Z(h),
\label{sa2}
\end{equation}
where $\beta^{-1}=T$, where $T$ is the temperature of the system. This averaged free energy represents, in an 
Euclidean field theory, the connected vacuum-to-vacuum diagrams in the disordered system. For a general disorder 
probability distribution, using the disordered functional integral $Z(h)$ given by Eq. (\ref{8}), the distributional 
zeta-function, $\Phi(s)$, is defined as

\begin{equation}
 \Phi(s)=\int [dh]P(h)\frac{1}{Z(h)^{s}},
\label{pro1}
\end{equation}
for $s\in \mathbb{C}$, this function being defined in the region where the above integral converges. The above 
equation is a natural generalization of the families of zeta-functions 
\cite{riem,riem2,landau,fro,primes0,primes,voros}. The average free energy can be written as 

\begin{equation}
 F=(d/ds)\Phi(s)|_{s=0^{+}}, \,\,\,\,\,\,\,\,\,\, \Re(s) \geq 0,  
\end{equation}
where one defines the complex exponential $n^{-s}=\exp(-s\log n)$, with $\log n\in\mathbb{R}$. Using analytic tools, 
the average free energy can be represented as

\begin{equation}
 F=\frac{1}{\beta}\Biggl[\sum_{k=1}^\infty \frac{(-1)^{k}a^{k}}{k}\,\mathbb{E}\,[Z^{\,k}]+\ln(a)+\gamma-R(a)\Biggr]
\label{m23e}
\end{equation}
where $a$ is a dimensionless arbitrary constant, $\gamma$ is the Euler-Mascheroni constant, and, for large $a$, 
$|R(a)|$ is quite small, therefore, the dominant contribution to the average free energy is given by the replica 
partition functions of the model. For simplicity we write 
$\mathbb{E}\left[Z(h)^{k}\right]\equiv\mathbb{E}\left[Z^{k}\right]$. Note that a $\frac{1}{k!}$ factor was absorbed 
in $\mathbb{E}\,[Z^{\,k}]$. To proceed, we  assume that the probability distribution of the disorder is written as 
$[dh]\,P(h)$, where

\begin{equation}
 P(h)=p\,\exp\Biggl(-\frac{1}{2\,\sigma}\int\,d^{d}x(h(x))^{2}\Biggr).
\label{dis2}
\end{equation}
The quantity $\sigma$ is a positive parameter associated with the disorder and $p$ is a normalization constant. In 
this case we have a delta correlated disorder field, i.e., $\mathbb{E}[{h(x)h(y)}]=\sigma\delta^{d}(x-y)$. As it was 
stressed by many authors, it is important to clarify the behavior of the model for small values of $\sigma$. After 
integrating over the disorder we get that each replica partition function $\mathbb{E}\,[Z^{\,k}]$ can be written as

\begin{equation}
 \mathbb{E}\,[Z^{\,k}]=
      \frac{1}{k!}\int\,\prod_{i=1}^{k}[d\varphi_{i}]\,\exp\Bigl(-H_{\textrm{eff}}(\varphi_{i})\Bigr),
\label{aa11}
\end{equation}
where the effective Hamiltonian $H_{\textrm{eff}}(\varphi_{i})$ describing the field theory with $k$-replica field 
components is given by

\begin{align}
 H_{\textrm{eff}}(\varphi_{i})&=
 \int d^{\,d}x\Biggl[\sum_{i=1}^{k}\biggl(\frac{1}{2}\varphi_{i}(x)\bigl(-\Delta+m_{0}^{2}\bigr)\varphi_{i}(x)
   \nonumber\\
&+\frac{\lambda_{0}}{4!}\varphi_{i}^{4}(x)\biggr)
     -\frac{\sigma}{2}\sum_{i,j=1}^{k}\varphi_{i}(x)\varphi_{j}(x)\Biggr].
\label{Seff1}
\end{align}

In the original Landau mean-field theory to discuss second-order phase transitions, an expansion for the free energy 
near the critical temperature as a power series of the order parameter is introduced. It is important to keep in mind 
that in the framework discussed by us the same idea is introduced. Nevertheless, by the presence of the disorder 
field, instead of a series in the order parameter we get a series in the replica partition functions of the model to 
define the averaged free energy.

The mean-field theory corresponds to a saddle-point approximation in each replica partition function. A perturbative 
approach gives us the fluctuation corrections to mean-field theory. Hence, to implement a perturbative scheme, it is 
necessary to investigate fluctuations around the mean-field equations. From each replica field theory, let us 
investigate the solutions of the saddle-point equations which are given by

\begin{equation}
 \Bigl(-\Delta\,+m_{0}^{2}\Bigr)\varphi_{i}(x)
                     +\frac{\lambda}{3!}\varphi^{3}_{i}(x)=\sigma\sum_{j=1}^{k}\varphi_{j}(x).
\label{sp}
\end{equation}
Imposing the replica symmetric ansatz, $i.e.$, $\varphi_{i}(x)=\varphi(x)$, the saddle-point equation, in each 
replica partition function, reads 

\begin{equation}
 \Bigl(-\Delta\,+m_{0}^{2}-k\sigma\Bigr)\varphi(x)+\frac{\lambda_{0}}{3!}\varphi^{3}(x)=0.
\label{sp1}
\end{equation}

At this stage it is easy to understand why the original replica method has problems, at least in this model. In this 
method, the average free energy is obtained using the formula 

\begin{equation}
 \mathbb{E}\,[{\,\ln Z(h)}]=\lim_{n\rightarrow 0}\frac{\partial}{\partial n}\mathbb{E}\,[ Z(h)^{n}].
\end{equation}

The $n\rightarrow 0$ limit in Eqs. (\ref{aa11}), (\ref{Seff1}) is translated to a field theory with the dimension of 
the order parameter going to zero. Therefore, we would like to briefly discuss the limit $n\rightarrow 0$ in the 
$O(n)$ Landau-Ginzburg model. It is well known that the self-avoiding random walk can be used as a mathematical model 
for polymers chains, where effects of excluded volume must be modeled \cite{fisher,dhar}. Since it represents a non-
Markovian stochastic process, there are many open questions in the literature, as, for instance, how many walks there 
are between two points. In the case of the self-avoiding random walk problem, the probability of finding the particle 
at $y$ at time $t$ if the particle was released in point $x$ at $t=0$, is a sum of diagrams that are exactly those 
for the correlation function of the $O(n)$ Landau-Ginzburg model for $n\rightarrow 0$. 

In the original replica method although one work with a replica field theory where the number of replicas must go to 
zero, the situation is quite different from the above discussed cases. The average free energy  involves derivation 
of the integer moments of the partition function. One consequence of this fact is that using the simplest possible 
replica symmetric ansatz in each replica partition function reduce the equations to the saddle-point equations of 
systems without disorder. Therefore, the replica symmetry breaking is introduced as a necessary condition to recover 
information from the disorder field in the theory. 

Using the distributional zeta-function method we can go further, since we have obtained analytic expression for the 
average free energy that does not involve derivation of such integer moments. Notice that, in principle, we have to 
consider all terms in Eq. (\ref{m23e}), since all values of $k$ are allowed. However, we have a constraint as the 
squared mass, $m_{0}^{2}-k\sigma$, must be positive definite to describe a well-defined physical theory. In this 
case, one has a critical value of $k$, namely, $k_{c}=\left\lfloor m_{0}^{2}/\sigma\right\rfloor$, above which one 
would obtain a negative squared mass, where $\left\lfloor x \right\rfloor$  means the integer part of $x$. For 
$k<k_{c}$, the replica fields fluctuate around the zero value. For $k>k_{c}$, we have to shift these replica fields 
since the zero value is not a stable equilibrium state. The last situation represents a spontaneous symmetry breaking 
mechanism. 

In the framework of distributional zeta-function method, defining 
$v=\left(\frac{6(\sigma N-m_{0}^{2})}{\lambda_{0}}\right)^{1/2}$, the simplest choice of the replica space is 
given by

\begin{equation}
\begin{cases}
\varphi_{i}^{(l)}(x)=\varphi(x) \;\;\; \hbox{for $l=1,\cdots,k_{c}$ and $i=1,\cdots,l$} \\
%\varphi_{i}^{(l)}(x)=\varphi_{i}(x)=\varphi(x)\;\;\hbox{for $l=k_{c}+1,\cdots,N$ and $i=1,\cdots,k_{c}$} \\
\varphi_{i}^{(l)}(x)=\phi(x)+v \;\;\hbox{for $l=k_{c}+1,\cdots,N$ and $i=1,\cdots,l$} \\
\varphi_{i}^{(l)}(x)=0 \quad \,\,\,\,\,\hbox{for $l>N$.}
\end{cases} 
\label{RSB2}
\end{equation}
Notice that we find a positive squared mass with self-interactions terms $\phi(x)^{3}$ and $\phi(x)^{4}$. From Eq. 
(\ref{RSB2}) and for $a$ and $N$  very large, the average free energy can be written as

\begin{equation}
 F=\frac{1}{\beta}\sum_{k=1}^{N} \frac{(-1)^{k}a^{k}}{k}\,\mathbb{E}\,[Z^{\,k}], 
\label{KMenergy}
\end{equation}
which has its leading term for $k=N$. Therefore, in the large-$N$ limit, the expression for disorder-averaged free 
energy is reduced to the contribution of only one replica partition function, consisting in a large $N$-component 
replica fields. In the context of a large-$N$ scenario, we introduce two 't Hooft couplings, namely, $f_{0}=\sigma N$ 
and $g_{0}=\lambda_{0} N$. These parameters are finite in $N\rightarrow \infty$ although $\lambda\rightarrow 0$ and 
$\sigma\rightarrow 0$. 

\section{Replica instantons-like solutions in the disordered system} \label{sec:replicainstantons}
\quad

The mean-field approach approach is used to analyze the phase diagram of our model. First, we consider that 
$m_{0}^{2}$ is a regular function of temperature. This situation is more complex than in a ordered system.  We find 
three regions of interest. The first occurs for $m_{0}^{2}\geq\sigma N$. In this case, all the replica fields 
oscillate around $\varphi=0$, the trivial vaccum. For $a\gg N$,  a very large $N$ limit is represented by only one 
replica partition function with $N$ ($N$ even) replica fields $\phi_{i}$. The $N$ replica fields has the symmetry 
$[\mathbb{Z}_{2}\times\mathbb{Z}_{2}\cdots\times\mathbb{Z}_{2}]$. There is also a critical temperature $T^{(1)}_{c}$, 
where $m_{0}^{2}=N\sigma$. The $[\mathbb{Z}_{2}\times\mathbb{Z}_{2}\cdots\times\mathbb{Z}_{2}]$ symmetry is broken 
below $T^{(1)}_{c}$. For the second region, $\sigma \leq m_{0}^{2}<\sigma N$, replica fields in some partition 
functions oscillates around the trivial vaccum, whereas fields in other replica partition functions now oscillates 
around the non-trivial vacuum. We are not interested in these ranges of $m_{0}^{2}$, for more details see Ref. 
\cite{zarro}. For $m_{0}^{2}<\sigma$, all the replica fields in each replica partition functions are oscillating 
around the non-trivial vacuum. In this case, for $a \gg N$ and for a very large-$N$ limit ($N$ even), the average 
free energy reads

\begin{equation}
 F=\frac{1}{\beta}\,\mathbb{E}\,[{Z^{\,N}}], \label{LGEnergyCase3}
\end{equation}
where $a$ is absorbed in normalization of the functional integration and $\mathbb{E}[{Z^{\,N}}]$ is written as

\begin{equation} \label{E2(Zk)}
\mathbb{E}[{Z^{\,N}}]=\frac{1}{N!}\int\,\prod_{i=1}^{N}\left[d\phi_{j}\right]\,\exp\biggl(-
H_{\textrm{eff}}\left(\phi_{j}\right)\biggr),
\end{equation}
and the effective Hamiltonian $H_{\textrm{eff}}(\phi_{i})$ is given by

\begin{align}
H_{\textrm{eff}}(\phi_{i})=&\int d^{\,d}x\Biggl[\sum_{i=1}^{N}\biggl(\frac{1}{2}
\phi_{i}(x)\Bigl(-\Delta+3f_{0}-2m_{0}^{2}\Bigr)\phi_{i}(x)\nonumber\\
&\,+\Bigl(\frac{f_{0}g_{0}}{3!N}\Bigr)^{\frac{1}{2}}\left(1-\frac{m_{0}^{2}}{f_{0}}\right)^{\frac{1}
{2}}\phi_{i}^{3}(x)+\frac{g_{0}}{4!N}\phi_{i}^{4}(x)\biggr)\nonumber \\
&-\frac{f_{0}}{2N}\sum_{i,j=1}^{N}\phi_{i}(x)\phi_{j}(x)\Biggr].
\label{Seff(2)}
\end{align}
Notice that the symmetry $[\mathbb{Z}_{2}\times\mathbb{Z}_{2}\cdots\times\mathbb{Z}_{2}]$ for $N$ replica fields is 
broken. A relevant question in the random field Ising model concerns the existence of an upper critical dimension, 
which, above it, the mean field approximation is exact. Since we have a cubic term in the action, the upper critical 
dimension is obtained from the relation $\frac{3}{2}(d-2)=d$, where the coupling constant becomes dimensionless, 
therefore the critical dimension is $d=6$. This result was discussed by Imry and Ma \cite{ma} and more recently in 
Ref. \cite{Ahrens}. 

Our fundamental result is the following. To describe critical phenomena for systems without disorder it is introduced 
an order parameter that describes second-order phase transition where for low temperatures a state of reduced 
symmetry appears. In the disordered system the order parameter is now a $N$-vector field. Our aim is to describe 
bubble nucleation in the disordered model at low temperatures. A representation similar to the strong-coupling 
expansion in field theory \cite{SKD,RMDHS,CBFC,sce} or the linked cluster expansion 
\cite{lusher1,lusher2,lusher3,reisz1,reisz2} can be used to represent a replica field theory. Rather than the usual 
case, which relies upon a gradient-free action, now the replicas become connected after applying a functional 
differential operator on a well-defined replica partition function. Here we would like to stress that the use of the 
linked cluster expansion in the Ising model was introduced in the literature by Englert \cite{Englert}.

To proceed,  an external source $\mathcal{J}_{i}(x)$ in replica space linearly coupled with each replica is 
introduced. Defining $R(x-y)=\sigma\delta^{d}(x-y)$, each replica partition function, $\mathbb{E}\,
[Z^{N}]=\mathcal{Z}(\mathcal{J})$, is written as a functional differential operator applied on $Q_{0}(\mathcal{J})$. 
Hence

\begin{align}
 &\mathcal{Z}(J)=\nonumber \\
 &\exp\biggr[-\frac{1}{2}\sum_{i,j=1}^{N}\int d^{\,d}x\,d^{\,d}y\frac{\delta}{\delta 
 \mathcal{J}_{i}(x)}R\frac{\delta}{\delta \mathcal{J}_{j}(y)}\biggr]Q_{0}(J).
\label{instanton1}
\end{align}
In the above equation, $Q_{0}(\mathcal{J})$, a modified replica partition function, is written as
 
\begin{equation}
  Q_{0}(\mathcal{J})=\frac{1}{N!}\int\prod_{j=1}^{N}[d\phi_{j}]\,\exp\biggl(-H_{\textrm{eff}}^{(0)}(\phi_{j},
  \mathcal{J}_{i})\biggr),
\label{instanton2}
\end{equation}
where $H_{\textrm{eff}}^{(0)}(\phi_{i},\mathcal{J}_{i})$ is given by

\begin{align}
&H_{\textrm{eff}}^{(0)}(\phi_{i},\mathcal{J}_{i})=\int d^{\,d}x \sum_{i=1}^{N}\Biggl[\frac{1}{2}\phi_{i}
(x)\Bigl(-\Delta+3f_{0}-2m_{0}^{2}\Bigr)\phi_{i}(x)\nonumber\\
&\Bigl(\frac{f_{0}g_{0}}{3!N}\Bigr)^{\frac{1}{2}}\left(1-\frac{m_{0}^{2}}{f_{0}}\right)^{\frac{1}{2}}\phi_{i}^{3}
(x)+\frac{g_{0}}{4!N}\phi_{i}^{4}(x)+\mathcal{J}_{i}(x)\phi_{i}(x)\Biggr].\nonumber \\
\label{instantonN}
\end{align}
Notice that the above equation does not contain interaction terms between replica fields. It is important to notice 
that Eqs. (\ref{instanton2}) and (\ref{instantonN}) fixes all ultraviolet divergences of our model that can be 
regularized by standard analytic regularization procedures \cite{bol,analit1,analit2,analit3,analit5}. 
The main idea is that in the $\epsilon=(4-d)$ expansion all the primitively divergent correlation functions contain 
poles. The principal part of the Laurent expansion defines the counterterms that we have to introduce to cancel such 
polar contributions. Introducing the renormalization constants $Z_{\varphi}$, $Z_{\lambda}$  and $Z_{m}$ the theory 
becomes finite. This pertubative expansion program with the regularization and renormalization procedures can be 
straightforwardly implemented. However, we will not follow it further in this analysis. Instead, we will study the 
vacuum structure of the first factor of Eq. (\ref{instanton1}), 
$i.e.$, $\mathcal{Z}(\mathcal{J})= Q_{0}(\mathcal{J})$. It is possible to define the generating functional of 
connected correlation functions $\mathcal W(\mathcal{J})=\ln \mathcal{Z}(\mathcal{J})$. For simplicity we assume that 
we have one replica field. The generating functional of one-particle irreducible correlations (vertex functions), 
$\Gamma[\overline{\phi}]$, is gotten by taking the Legendre transform of $\mathcal{W}(\mathcal{J})$ \cite{amit}

\begin{equation} 
 \Gamma[\overline{\phi}]+\mathcal{W}(\mathcal{J})=\int d^{\,d}x\biggl(\mathcal{J}(x)\overline{\phi}(x)\biggr),
\end{equation}
where 

\begin{equation}
 \overline{\phi}(x)=\left.\dfrac{\delta \mathcal{W}(\mathcal{J}) }{\delta \mathcal{J}}\right|_{\mathcal{J}=0}. 
\end{equation}

Now, we assume that the field $\overline{\phi}(x)=\phi$, is uniform. In this case, we can write the effective 
potential, $V(\phi)$, as

\begin{equation} 
 \Gamma[\phi]=\int\,d^{d}x \,V(\phi),
\end{equation}
where $V(\phi)$  takes into account the fluctuations in the model. From above discussion it is possible to write the 
tree-level effective potential for each replica field in the leading replica partition function. We have 
$V_{tree}(\phi)=U(\phi)$ where

\begin{equation}
  U(\phi)=\frac{1}{2}(3f_{0}-2m_{0}^{2})\phi^{2}+\frac{\lambda_{0} v}{3!}\phi^{3}+\frac{\lambda_{0}}{4!}\phi^{4},
\label{potential}
\end{equation}
where $v=\sqrt{6(f_{0}-m_{0}^{2})/\lambda_{0}}$ and the replica symmetric ansatz was evoked. The false and the true 
vacuum states $\phi_{(\pm)}$ can be obtained

\begin{equation}
 \phi_{(\pm)}=-\frac{3v}{2}\pm3\sqrt{-\frac{f_{0}}{2\lambda_{0}}-\frac{m^{2}_{0}}{6\lambda_{0}}}.
\label{raises}
\end{equation} 
Therefore, we get the following interesting result: \emph{there are instantons-like solutions in our model}. The 
first term in the series representation for the functional differential operator is the diluted instanton  
approximation, i.e., $N$ non-interacting instantons-like solutions. For $f_{0}>m_{0}^{2}>-3f_{0}$, the system 
develops a spontaneous symmetry breaking in the leading replica partition function. In this case, all $N$ instantons-
like solutions are complex. On the other hand, for $m_{0}^{2}<-3f_{0}$ we get a similar situation as before, however 
all the instantons-like solutions are real. 

Vacuum transition in this theory with $N$ replicas can be described in the following way. Lowering the temperature 
each replica field has two non-degenerate vacuum states. The transition from the false vacuum to the true one will 
nucleate bubbles of the true vacuum. This first-order phase transition, in the low temperature limit, was 
investigated in Refs. \cite{aharonytricritical,aharonyetal}. The crucial question here is which tools we can use to 
describe the nucleation of bubbles. 

\section{Bubble nucleation and the fermionic Dicke model}\label{sec:effectivemodel}
\quad

In this section, we introduce a quite simple model to study the collective nucleation of bubbles in the disordered 
system. Our aim is to transform the original problem substituting by one that is technically treatable where the 
physical essence of the original problem is maintained. Let us remind the reader that one fundamental problem in 
quantum optics is the description of spontaneous emission of atoms \cite{Einstein,Agarwal,Fonda}. In fluorescence 
situation, in the decay by spontaneous emission the atoms tend to decay independently. However, other regime also 
happens when the atoms act together. Superradiance is exactly this collective behavior when $N$ excited atoms in a 
cavity or in the free space where they are close together, with some characteristic length, decay spontaneously 
\cite{andreev,gross}. The Dicke model was introduced to describe such collective behavior 
\cite{Dicke,TBrandes,Barry}. In this model it is assumed that the system is composed by an ensemble of two-level 
atoms, all of them in the excited state initially. Furthermore one assume that the two-level atoms are trapped in a 
high-Q cavity, then effectively one single mode in the countable infinitely set of field modes trapped by the cavity
interact with the atoms. Other possibility is to assume that the two-level atoms interact with the free space 
continuum of field modes, but all the atoms are confined in a region with a characteristic length small compare with 
the wavelength of the resonant field mode. Both situations can describe a collective effect of emission, the 
superradiance, although irreversibility occurs only in the second situation, since the high-Q cavity makes the first 
situation time-invertible. In conclusion, this spin-boson model, even in the case of a single mode, is able to 
describes a phase transition from the fluorescent to superradiant phase, characterized by the fact that atoms in 
quite special conditions behaves cooperatively. They start to radiate spontaneously much faster and strongly than the 
emission of independent atoms. 

From the multimode Dicke model, with spatially varying coupling between the two-level atoms and the bosonic modes, a 
spin-glass behavior is obtained after integrating out the bosonic field \cite{Sarang, Sachev, Rotondo1, Rotondo2}.  
What firstly comes to mind is the feasibility of the reverse situation, $i.e.$, starting from the random field 
Landau-Ginzburg model, a particular disordered statistical field theory model defined in the continuum, to use the 
Dicke model to describe the phase transitions of the system. 

Let us start, discussing first the decay of one replica field from the false to the true vacuum state. Suppose that 
each replica field $\phi^{i}(x)$ is in the metastable state $\phi_{(+)}^{i}$. Let us assume that the free energy gap 
per unit volume between the metastable state  $\phi^{i}_{(+)}$ and the state $\phi^{i}_{(-)}$ is $\omega_{i}$. With 
the bubble formation of radius $R_{i}$ the free energy decreases by $\frac{4\pi}{3} R^{3}_{i}\omega_{i}$ inside the 
bubble. The interface makes the free energy increases by $4\pi R_{i}^{2}\eta_{i}$ where $\eta_{i}$ is the interface 
free energy per unit area for each replica field. The contribution for each replica field to the free energy 
$\Delta F_{i}$ is $4\pi R_{i}^{2}\eta_{i}-\frac{4\pi}{3} R^{3}_{i}\omega_{i}$. There is a critical radius $R_{c}$ 
where for $R>R_{c}$ the nucleation of bubbles occurs.  For finite temperature we have thermal nucleation of bubbles. 
In the case where $\beta\rightarrow \infty$ there is a quantum nucleation of bubbles. There is a standard  procedure 
to find the decay rate in a Euclidean scalar theory  \cite{SColeman1, SColeman2,Flores}. This formalism is not able 
to describe the collective behavior, i.e., the nucleation of $N$ bubbles. Since we would like to describe the 
nucleation of $N$ bubbles, we discuss here an alternative approach where the description of a cooperative behavior of 
two-level systems was presented.

Going back to the disordered model, lowering the temperature, each replica field has two non-degenerate vacuum 
states. The transition from the false vacuum to the true one will nucleate bubbles of the true vacuum. Our aim is to 
obtain an collective effective model to deal with a gas of $N$ real interacting instantons-like solutions 
(see, $e.g.$ Eq. (\ref{Seff(2)})). We claim that the qualitative features of the disordered system at very low 
temperatures can be described by the generalized Dicke model with only one single bosonic mode.   
In the Dicke model there is a mean-field type phase transition with a critical temperature below which the system is 
in a superradiant state. Some seminal papers discussing the phase transition in such model are Refs. 
\cite{Hepp1,Hepp2,Wang,Hioe}.  

Following Ref. \cite{Leggett}, it is possible to represent $N$ structures with a false  and true vacuum by $N$ two-
level systems. Referring to Eqs. (\ref{instanton1}), (\ref{instanton2}) and (\ref{instantonN}), we are modeling the 
effect of considering more terms of the series, $i.e.$, going beyond the diluted instanton approximation, as a 
bosonic mode interacting with all the two-level systems. The effective bosonic mode was introduced to play a two-fold 
role: is an effective mode that allows the interactions between the two-level systems and also to make the decay 
$\phi^{i}_{(+)}\rightarrow \phi^{i}_{(-)}$, possible. Note that we have actually an open system. In conclusion, the 
situations where nucleation of bubbles occurs, decreasing the free energy of the system will be substituted by an 
effective model. It is important to point out that we have assumed that going beyond the diluted instanton 
approximation, the vacuum structure associated to each replica field is not modified. If the inclusion of more terms 
of the series defined by Eq. (\ref{instanton1}) increase number of false vacuum states for each replica field,  it is 
necessary to generalize the Dicke model using intermediate statistics \cite{Greenberg1,Greenberg2}.  

In order to achieve the effective description discussed above, let us introduce, following Popov and Fedotov, the 
fermionic generalized Dicke model. See also Refs. \cite{Aparicio1,Aparicio2,Aparicio3,Aparicio4}. To proceed, let us 
define an auxiliary model to be called the fermionic full Dicke model in terms of fermionic raising and lowering 
operators $\alpha^{\dagger}_{i}$, $\alpha_{i}$, $\beta^{\dagger}_{i}$ and $\beta_{i}$, that satisfy the anti-
commutator relations $\alpha_{i}\alpha^{\dagger}_{j}+\alpha^{\dagger}_{j}\alpha_{i}=\delta_{ij}$ and 
$\beta_{i}\beta^{\dagger}_{j}+\beta^{\dagger}_{j}\beta_{i}=\delta_{ij}$. We can also define the following bilinear 
combination of fermionic operators, 
$\alpha^{\dagger}_{i}\alpha_{i} -\beta^{\dagger}_{i}\beta_{i}$, $\alpha^{\dagger}_{i}\beta_{i}$ and 
$\beta^{\dagger}_{i}\alpha_{i}$ which obey the same commutation relations as the pseudo-spin operators 
$\sigma^z_{(\,i)}$, $\sigma^+_{(\,i)}$ and $\sigma^-_{(\,i)}$.

\begin{equation}
 \sigma_{i}^{z}\longrightarrow \alpha_{i}^{\dagger}\alpha_{i}-\beta_{i}^{\dagger}\beta_{i}\, , \label{34}
\end{equation}

\begin{equation}
 \sigma_{i}^{+}\longrightarrow \alpha_{i}^{\dagger}\beta_{i}\, ,
\label{35}
\end{equation}
and finally

\begin{equation}
 \sigma_{i}^{-}\longrightarrow \beta_{i}^{\dagger}\alpha_{i}\, .
\label{36}
\end{equation}

The Hamiltonian $ H_F$ of the auxiliary fermionic full Dicke model is

\begin{equation}
  H_F=\frac{\Omega}{2}\,\sum_{i=1}^{N}(\alpha_{i}^{\dagger}\alpha_{i}-\beta_{i}^{\dagger}\beta_{i})
  +\,\omega_0\,b^{\dagger}\,b\,+\frac{g_1}{\sqrt{N}}\sum_{i=1}^{N} 
  \left(b\,\alpha_{i}^{\dagger}\beta_{i}+b^{\dagger}\,\beta_{i}^{\dagger}\alpha_{i}\right)
  \,+\frac{g_2}{\sqrt{N}}\sum_{i=1}^{N} 
  \left(b\,\beta_{i}^{\dagger}\alpha_{i}+b^{\dagger}\,\alpha_{i}^{\dagger}\beta_{i}\right),
\label{HFdipole}
\end{equation}
where $\Omega$ is a known function of $m_{0}$, $\lambda_{0}$ and $f_{0}$. It is related to the energy gap between the 
false and the true vacuum for each replica field. See Eq. (\ref{raises}). On the other hand,  $\omega_{0}$, $g_{1}$ 
and $g_{2}$ are phenomenological quantities that are related to the physical parameters $m_{0}$, $\lambda_{0}$ and 
$f_{0}$ of the disordered model. In this situation, the Euclidean action $S$ associated to the fermionic Dicke model 
is given by

\begin{equation}
 S=\int_0^{\beta} d\tau \left(b^*(\tau)\,\partial_{\tau}b(\tau)+ \sum_{i=1}^{N}
 \Bigl(\alpha^*_i(\tau)\,\partial_{\tau}\alpha_i(\tau)
 +\beta^*_i (\tau)\,\partial_{\tau}\beta_i(\tau)\Bigr)\right) -\int_0^{\beta}d\tau H_{F}(\tau)\,,
\label{66}
\end{equation}
where the Hamiltonian density $H_{F}(x)$ is obtained from Eq. (\ref{HFdipole}). In order to define the partition 
function, the functional integrals have to be done in the space of complex functions $b^*(\tau)$ and $b(\tau)$ and 
Grassmann variables $\alpha_i^*(\tau)$, $\alpha_i(\tau)$, $\beta_i^*(\tau)$ and $\beta_i(\tau)$. Since we use thermal 
equilibrium boundary conditions in the Euclidean time, the integration variables obey periodic boundary conditions 
for the Bose field, $i.e.$, $b(0)=b(\beta)$ and anti-periodic boundary conditions for Grassmann variables, i.e., 
$\alpha_i(\beta)=-\alpha_i(0)$ and $ \beta_i(\beta)=-\beta_i(0)$ \cite{kubo,martin}.  

To proceed, let us define the formal quotient of two functional integrals, i.e., the partition function of the 
generalized fermionic Dicke model and the partition function of the free fermionic Dicke model. Therefore we are 
interested in calculating the following quantity

\begin{equation}
 \frac{Z_{F}}{Z_{F_{0}}}=\frac{\int [d\eta]\,e^{\,S}}{\int[d\eta]\,e^{\,S_{0}}}\, , \label{65}
\end{equation}
where $S$ is the Euclidean action of the generalized fermionic Dicke model given by Eq. (\ref{66}), $S_0$ is the free 
Euclidean action for the free single bosonic mode and the free two-level systems. In the above equation $[d\eta]$ is 
the standard functional measure for the fermionic and bosonic degrees of freedom. The free action for the single mode 
bosonic field $S_{0}(b)$ is given by

\begin{equation}
S_{0}(b, b^{*}) = \int_{0}^{\beta} d\tau \biggl(b^{*}(\tau)\frac{\partial b(\tau)}{\partial \tau} 
 -\omega_{0}\,b^{*}(\tau)b(\tau)\biggr)\, . \label{67}
\end{equation}

Then we can write the action $S$ of the generalized fermionic Dicke model, given by Eq. (\ref{66}), using the free 
action for the single mode bosonic field $S_{0}(b, b^{*})$, given by Eq. (\ref{67}), plus an additional term that can 
be expressed in a matrix form. For more details see the Refs. \cite{Aparicio1,Aparicio2}. Performing straightforward 
calculations it is possible to show that the critical temperature $T_{c}$ where $T^{-1}=\beta$, is

\begin{equation}
\beta_{c}=\frac{2}{\Omega}\text{arctanh}\left[\frac{\omega_{0}\,\Omega}{(g_1+g_2)^2}\right]\,.
\label{eqfase11}
\end{equation}
Notice that it is possible to have a quantum phase transition when $\omega_{0}\,\Omega=(g_1+g_2)^2$. 
The experimental realization of the Dicke superradiance in cold atoms in optical cavities was presented in Ref. 
\cite{bauman}.

In the disordered system, this situation discussed above corresponds to the quantum nucleation of bubbles. 
We would like to stress that this scenario, where these bubble nucleations are a collective effect in the system, 
is a oversimplification of the exact full model. 
At this point we would like to comment the similarities between these two physical systems, the $N$ two-level systems 
trapped in a cavity and the random field Landau-Ginzburg model. In the first case, the ensemble of two-level atoms 
interacts effectively with one bosonic field mode present in the cavity. There are strong correlations between the 
two-level systems. In the disordered Landau-Ginzburg model the Gaussian disorder is able to make the same effect of 
the cavity. All the replicas are strongly correlated. See Eq. (\ref{Seff(2)}). All the replicas are under the effect 
of the background generated by the other replicas. 
  
\section{Conclusions}\label{sec:conclusions}
\quad

In this work we discuss the phase transitions in the continuous version of the $d$-dimensional random field Ising 
model. First we adopt the general strategy to average over the disordered free energy. Recently it was proposed a  
new method to average the disorder dependent free energy in systems defined in the continuum. Using this technique, 
the free energy is represented as a series of the replica partition functions of the system. The structure of the 
replica space was investigated using the saddle-point equations obtained from each replica field theory. We discuss 
the presence of a spontaneous symmetry breaking mechanism in some replica partition functions. For very low 
temperatures there are $N$ replica instantons-like solutions (real or complex) in this model. For the case of real 
instantons-like solutions, each replica field has two non-degenerate vacua. The transition from the false vacuum to 
the true one for each replica field corresponds to  the nucleation of bubble of the true vacuum. 

As we discussed, it is possible to obtain a spin-glass behavior from the multimode Dicke model of quantum optics, 
integrating out the bosonic field. This spin-boson model describes a phase transition from the fluorescent to 
superradiant phase. We show that the reverse situation is also feasible. To describe the phase transition in the 
disordered statistical field theory model we use the one mode Dicke model. The similarities between these two 
physical systems, the $N$ two-level systems trapped in a cavity and the random field Landau-Ginzburg model are 
evident.

The ensemble of two-level atoms interact effectively with one bosonic field mode present in the cavity. This fact 
generates strong correlations between the two-level atoms. In the disordered Landau-Ginzburg model, the Gaussian 
disorder is able to make the same effect, since all the replicas are strongly correlated.  All the replicas are under 
the effect of the background generated by the other replicas. 

Using the formalism developed by Popov and Fedotov the critical temperature is found. 
This temperature can be characterized by a non-analytical behavior of the thermodynamic quantities as a function of 
the temperature. At this temperature the free energy of the system is non-analytic, and the system present a 
transition to the normal to the superradiant phase.

A crucial question is the size of the bubbles in the disordered model. 
In scalar models in field theory with compactification in one spatial direction, the mass can depend upon the 
periodicity length in the 
compact direction \cite{LHFord,Toms,Toms2,Denardo,ford}. This situation allow that topological effects play a role in 
the breaking and restoration of symmetries in different models.
We believe that using the formalism discussed in this section and the above discussed mechanism, it is possible to to 
predict the size of the nucleating bubbles. 

Another natural continuation of our investigations still using the distributional zeta-function method in disordered 
field theory models, consists in studying the nature of phase transitions in the disordered (random temperature) $d$-
dimensional Ising ferromagnet, which can be 
described by a statistical field theory model with quenched disorder, i.e., the $d$-dimensional random temperature 
Landau-Ginzburg model. 

As we discussed in Sec. \ref{intro}, two concepts that are of fundamental importance in disordered systems are 
respectively quenched disorder and frustration.  
The presence of frustration in some disordered systems, as for example the spin glasses suggests that there are
many different ground states in such systems. At low temperatures, in the spin-glass there are domains where the 
spins becomes frozen in space. This randomness in space that characterize the spin-glass phase corresponds to the 
fact that the free energy landscape  of the system has a multivalley structure.
Some authors discussed the possibility of a existence of this multivalley structure of the spin-glass phase in the 
random temperature Landau-Ginzburg model \cite{ma1, targus}.

Our aim is to investigate the possibility of found a multivalley structure in the average free energy of the random 
temperature Landau-Ginzburg model  using the distributional zeta-function approach. This subject is under 
investigation by the authors.
 
\section{Acknowledgments}

We would like to thank G. Krein, S. Queir\'{o}s G. Menezes and M. Aparicio Alcalde for useful discussions.
This paper was partially supported by Conselho Nacional de Desenvolvimento Cient\'ifico e Tecnol{\'o}gico (CNPq, 
Brazil).

\end{document}